\documentclass[aps,prb,showpacs,floatfix,superscriptaddress,twocolumn]{revtex4}
\usepackage{graphicx}

\begin{document}

\title{Electronic structure of nanoscale iron oxide particles \\ measured by scanning tunneling and photoelectron spectroscopies}

\author{M. Preisinger} \author{M. Krispin} \author{T. Rudolf} \author{S. Horn}
\affiliation{Institute of Physics, University of Augsburg, 86135
Augsburg, Germany}

\author{D. R. Strongin}
\affiliation{Department of Chemistry, Temple University,
Philadelphia, Pennsylvania 19122, USA}

\date{\today}

\begin{abstract}

We have investigated the electronic structure of nano-sized iron
oxide by scanning tunnelling microscopy (STM) and spectroscopy
(STS) as well as by photoelectron spectroscopy. Nano particles
were produced by thermal treatment of Ferritin molecules
containing a self-assembled core of iron oxide. Depending on the
thermal treatment we were able to prepare different phases of iron
oxide nanoparticles resembling $\gamma$-Fe$_{2}$O$_{3}$,
$\alpha$-Fe$_{2}$O$_{3}$, and a phase which apparently contains
both $\gamma$-Fe$_{2}$O$_{3}$ and $\alpha$-Fe$_{2}$O$_{3}$.
Changes to the electronic structure of these materials were
studied under reducing conditions. We show that the surface band
gap of the electronic excitation spectrum can differ from that of
bulk material and is dominated by surface effects.

\end{abstract}


\maketitle

\section{Introduction}
Nano-sized transition metal oxides particles can be expected to
show size dependent optical, magnetic and chemical properties with
possible applications in catalysis and magnetic and optical
devices. It has already been shown, that catalytic and magnetic
properties of nanoparticles might be quite different from bulk
materials. An example are nano-sized particles of the compound
Fe$_{2}$O$_{3}$, which exhibit
superparamagnetism\cite{Berkowitz68,Janzen03,Raming02}, and, at
the same time, enhanced catalytic properties compared to the bulk
material\cite{Li03}. The size dependence of the catalytic
properties can be associated with an influence of size on the
electronic structure of the material. This influence can be due to
the surface tension of the particle, resulting in an effective
pressure or to changes of the stoichiometry (and resulting changes
of the state of oxidation of Fe) of such particles, driven by  the
large surface to volume ratio.

Here we investigate the electronic structure of nano-sized iron
oxide by scanning tunnelling microscopy and spectroscopy as well
as by photoelectron spectroscopy and optical spectroscopy. To
address the question of stoichiometry and oxidation state the
effect of oxidizing and reducing atmosphere on the electronic
structure of the nano particles is also investigated. To produce
nano-sized iron oxide particles we utilized the ability of
Ferritin to self-assemble and construct a core of iron oxide.
Ferritin is the major cellular iron-storage protein and consists
of a spherical hollow shell composed of 24 polypeptide subunits
and is able to store iron as hydrated iron oxide in the internal
cavity\cite{biomin89,Mass93,Ironox96}. The inner and outer
diameters of the protein shell are about 8 and 12.5~nm,
respectively. The iron core shows a structure similar to that of
the mineral ferrihydrite
(5~Fe$_{2}$O$_{3}$~\textperiodcentered~9~H$_{2}$O). Hydrophilic
and hydrophobic channels penetrate the protein shell and provide
the means by which iron and other atoms can be accumulated within
or removed from the molecules. In previous work, Ferritin was used
to prepare several kinds of nanoparticles and
nanocomposites\cite{Wong96,Hosein04}. Furthermore, Ferritin
molecules inherently show photocatalytic activity\cite{Kim02}.

\section{Experimental}

Monolayers of Ferritin molecules were prepared by the
Langmuir-Schaefer technique\cite{LB90}. A Ferritin solution
(0.15~mg/ml horse spleen Ferritin from Sigma) in 10~mM NaCl with a
pH of 5.5 was filled in a commercially available Langmuir-Trough
(Nima Technology). A few micrograms of trimethyloctadecylammonium
bromide (Fluka) were solved in chloroform to form a concentration
of 1~mg/ml and was then spread over the Ferritin solution
subphase. After compression to a surface pressure of 30~mN/m and
an adsorption time of a few minutes the monolayer was transferred
to a freshly cleaved surface of a highly oriented pyrolytic
graphite (HOPG) substrate by the horizontal lifting method.
Further details of preparation of 2D arrays are described
elsewhere\cite{Johnson00,Furuno89}.

Prior to the introduction into the ultra high vacuum (UHV) chamber
all samples were annealed in air at different temperatures and for
different time intervals ($\sim$~575~K for one hour, $\sim$~675~K
for one hour or $\sim$~675~K for five hours) to eliminate the
protein shell and to further oxidize the iron core. After
introduction into the chamber the samples were annealed \textit{in
situ} under a 1$\times$10$^{-6}$~mbar oxygen atmosphere at the
respective temperature for a few minutes. Utilizing this process
we produced dense arrays of iron oxide nanoparticles. To vary the
oxidation state of the respective nanoparticles, we reduced the
sample within the UHV system by exposure to 1$\times10^{-6}$~mbar
of an H$_2$/Ar-atmosphere (25\%~H$_2$, 75\%~Ar) at $\sim$~675~K
for two hours.

Spectroscopic measurements were carried out in an Omicron UHV
system, equipped with a DAR400 X-ray radiation source
(Al~K$\alpha$ 1486.6~eV) and an AR65 electron analyzer for X-ray
photoelectron spectroscopy (XPS). Scanning tunnelling microscopy
(STM) and -spectroscopy (STS) measurements were performed using a
variable temperature scanning probe microscope (VT-SPM) within the
same UHV system. All STM images were recorded in the constant
current mode, using a tungsten tip at a sample bias of -1.5~V and
a current of 0.1~nA. The topographic images represent the height
z(x,y) of the tunnelling tip above the sample after a plane and
slope subtraction was performed. The STS measurements were carried
out placing the tunnelling tip above a single nanoparticle and
keeping the tip-sample separation fixed while recording the I(U)
characteristic. A Bruker fourier-transform infrared spectrometer
(FTIR) was used for optical spectroscopy. STM and STS measurements
were done on monolayers of nanoparticles, while XPS and optical
spectroscopy measurements were done on thicker samples in order to
improve the signal to noise ratio.  Such samples were prepared
from a small dried drop of Ferritin solution (1~mg/ml) annealed in
air at the respective temperatures.

\section{results}

\begin{figure}
\includegraphics[width=8.5cm]{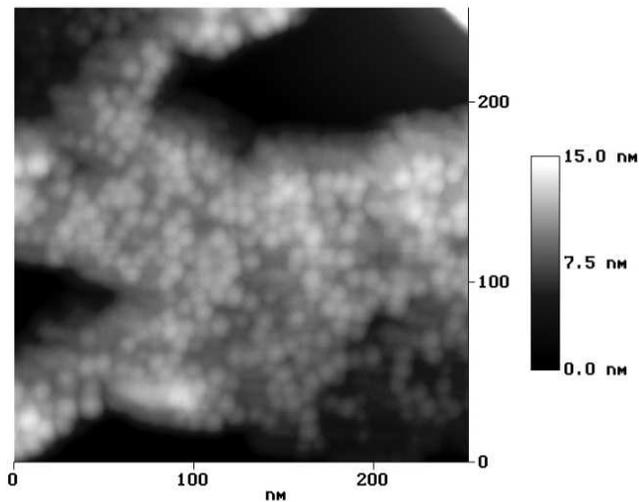}
\caption{\label{fig1}AFM image taken on a Ferritin monolayer
without annealing}
\end{figure}

Images of monolayers of Ferritin molecules under air
(Fig.~\ref{fig1}) were taken by Tapping Mode atomic force
microscopy (TM-AFM, utilizing a Nanoscope~IIIa, Digital
Instruments). The layer of Ferritin molecules is not completely
closed. The difference in height between the Ferritin layer and
the subjacent HOPG substrate of about 10~nm corresponds to the
thickness of a single Ferritin monolayer. Also the diameter of the
spheres observed is approximately 10~nm, characteristic of
Ferritin molecules with a dehydrated protein shell caused by
drying under air.

Depending on oxidation parameters, we obtain nanoparticles of
different iron oxide phases. X-ray diffraction (XRD) performed on
a thick layer of Ferritin annealed at $\sim$~675~K under air for
one hour (procedure~1) reveals rather broad reflections
characteristic of $\gamma$-Fe2O3 with some contributions of
$\alpha$-Fe$_{2}$O$_{3}$. In the following we will refer this
mixed phase as $\chi$-Fe$_{2}$O$_{3}$. Further annealing for five
hours (procedure~2) resulted in narrow reflections characteristic
of $\alpha$-Fe$_{2}$O$_{3}$. These results are consistent with
transmission electron microscopy (TEM) studies, which will be
described elsewhere\cite{Bremser04}. Annealing Ferritin molecules
at $\sim$~575~K in air (procedure~3) results in
$\gamma$-Fe$_{2}$O$_{3}$ nanoparticles, as confirmed by TEM.

\begin{figure}
\includegraphics[width=8.5cm]{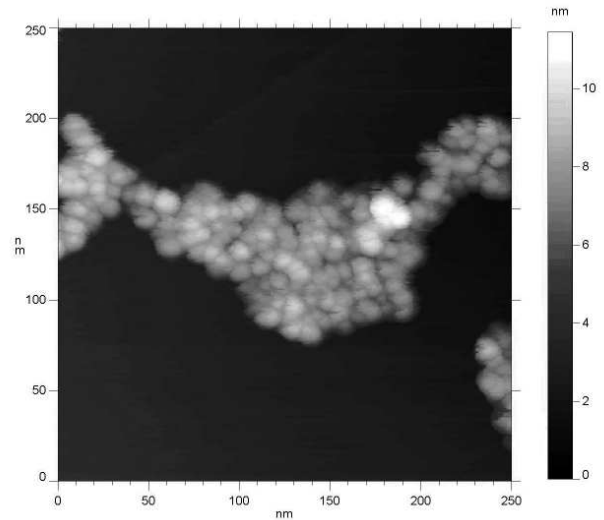}
\caption{\label{fig2}STM image of an oxidized Ferritin monolayer,
showing iron oxide nanoparticles}
\end{figure}

A STM image of an nanoparticle monolayer after annealing in air
(procdure~1) is shown in Figure~\ref{fig2}. Here we note that the
images of the nanoparticles of all three phases (not shown) look
alike. The image shows densely packed clusters of distinguishable
nanoparticles. After the annealing process the removal of the
protein shell results in a lower coverage of the HOPG surface. The
nanoparticle clusters consist of well defined particles of a size
of approximately 7~-~8~nm. Such a size is expected for the iron
oxide core of Ferritin. Clusters are found dominantly at steps
present on the HOPG substrate surface.

\begin{figure}
\includegraphics[width=8.5cm]{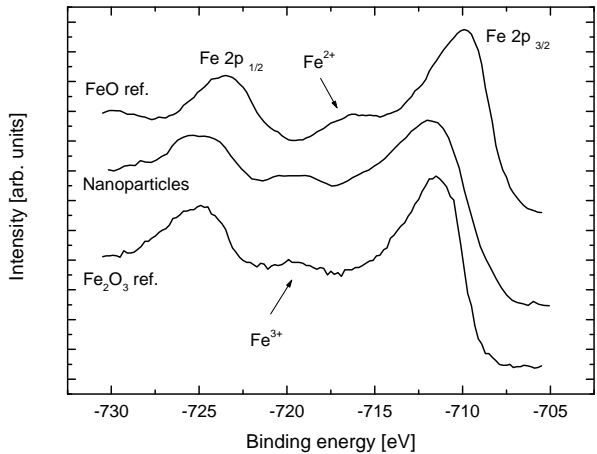}
\caption{\label{fig3}XPS spectra of oxidized Ferritin molecules
(Fe$_{2}$O$_{3}$ nanoparticles) and reference data of bulk-like
samples}
\end{figure}

To obtain information concerning the iron oxidation state we
performed XPS on the nanoparticles. Figure~\ref{fig3} displays the
Fe~2p XPS spectra characteristic of the oxidized nanoparticle
phases and, for comparison, the spectra of Fe$_{2}$O$_{3}$ and
FeO, taken from bulk-like material\cite{Weiss02,Oku99}. The latter
two spectra correspond to a Fe$^{3+}$ and a Fe$^{2+}$ state,
respectively. We did not observe any charging effects of the
nanoparticles within the instrumental width of about 0.2~eV. With
respect to the Fe$^{2+}$ spectrum, the Fe$^{3+}$ spectrum is
shifted to a higher binding energy by ~1.5~eV and shows a
characteristic satellite structure at a binding energy
E$_b$~=~719~eV.  The Fe$^{2+}$ spectrum shows a satellite
structure at E$_b$~=~716~eV. A Fe~2p$_{3/2}$ peak at 711.7~eV and
a satellite structure at about 719~eV in the spectrum taken from
the iron oxide nanoparticles is characteristic for a
Fe$^{3+}$-state, although the presence of a slight weight of
Fe$^{2+}$ can not be ruled out. Further oxidation of the particles
within the UHV system at an oxygen partial pressure of
1$\times10^{-6}$~mbar and a temperature of $\sim$~675~K does not
result in changes of the intensity of the characteristic
satellite, nor did it shift the spectrum to higher binding
energies. We, therefore, conclude that either i. the Fe- atoms of
the nanoparticles are already in the highest oxidation state
(Fe$_{2}$O$_{3}$), or ii. the Fe- atoms are not in the highest
oxidation state, but further oxidation of the nanoparticles is not
possible under the chosen conditions.

To investigate the electronic excitation spectrum of single
nanoparticles in the vicinity of the Fermi energy~E$_F$, STS
measurements were performed on the three different phases
identified by XRD and TEM ($\alpha$-Fe$_{2}$O$_{3}$,
$\gamma$-Fe$_{2}$O$_{3}$, $\chi$-Fe$_{2}$O$_{3}$). It is well
known, that tunnelling spectra obtained from nanocrystals may
depend on the tunnelling parameters and tunnelling configuration
due to e.g. charging\cite{Bakkers00,Katz01}, rendering the
determined gap width unreliable. Therefore we adjusted the
tunnelling parameters to perform the measurements at large
tunnelling tip-nanoparticle distance. The distance was increased
until a constant gap value was observed and charging effects did
not occur.

\begin{figure}
\includegraphics[width=8.5cm]{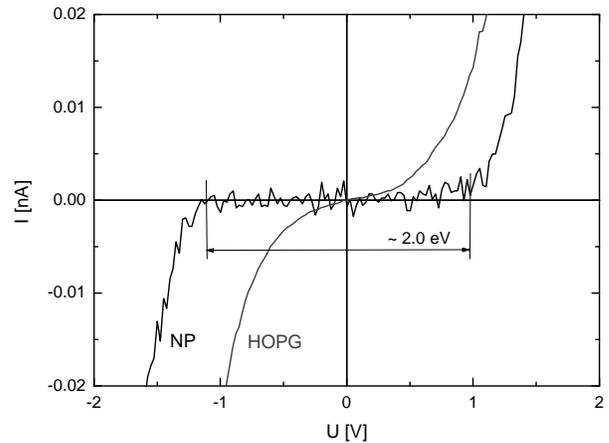}
\caption{\label{fig4}STS spectra of $\alpha$-Fe$_{2}$O$_{3}$
nanoparticles compared to HOPG substrate}
\end{figure}

\begin{figure}
\includegraphics[width=8.5cm]{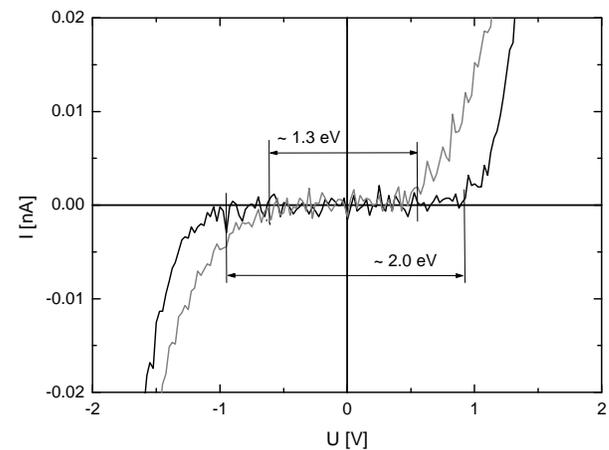}
\caption{\label{fig5}STS spectra of $\gamma$-Fe$_{2}$O$_{3}$
nanoparticles}
\end{figure}

\begin{figure}
\includegraphics[width=8.5cm]{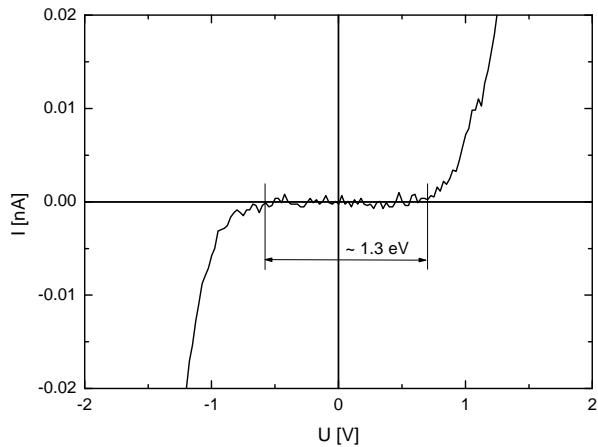}
\caption{\label{fig6}STS spectra of $\chi$-Fe$_{2}$O$_{3}$
nanoparticles}
\end{figure}

A typical tunnelling spectrum taken from a single
$\alpha$-Fe$_{2}$O$_{3}$ nanoparticle (NP) is plotted in
Figure~\ref{fig4}. For comparison, a spectrum taken from the HOPG
substrate is also shown. The I(U) - curves taken from the
nanoparticle clearly are characteristic for a gap energy of
$\Delta$~=~2.0~eV. The variation of the gap energy for spectra
taken from the same nanoparticle was within the error of the
measurement. Spectra were taken from of a few ten different
nanoparticles and the gap energy was found to vary between 1.8~eV
and 2.2~eV. This variation might reflect differences in particle
size, surface conditions and small variations of the tunnelling
junction parameters for the different measurements. The gap energy
of $\Delta$~=~2.0~eV corresponds to that measured on a bulk
$\alpha$-Fe$_{2}$O$_{3}$ sample
($\Delta$~=~2.2~eV)\cite{Ironox96}. Figure~\ref{fig5} displays
tunnelling spectra taken off $\gamma$-Fe$_{2}$O$_{3}$
nanoparticles. Here, spectra taken from different locations of a
single nanoparticle, show two different values of the gap
energies, $\Delta_{1}$~=~1.3~eV ($\pm$~0.2~eV) and
$\Delta_{2}$~=~2.0~eV ($\pm$~0.2~eV). We note, that the gap energy
of a bulk sample of $\gamma$-Fe$_{2}$O$_{3}$ shows a gap of 2.0~eV
\cite{Ironox96}. Finally, tunnelling spectra taken from
$\chi$-Fe$_{2}$O$_{3}$ phase nanoparticles show a band gap of 1.3~
eV ($\pm$~0.2~eV) (Fig.~\ref{fig6}), which does not change for
different locations on one nanoparticle. A gap value of 1.3~eV
does not correspond to the gap energy of any bulk iron oxide
phase\cite{Ironox96}.

\begin{figure}
\includegraphics[width=8.5cm]{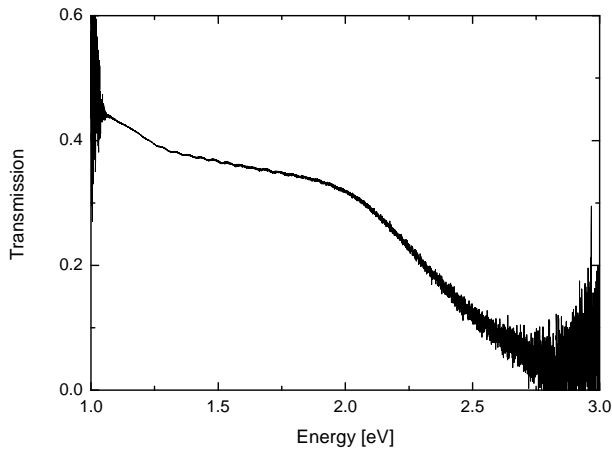}
\caption{\label{fig7}Optical transmission of
$\chi$-Fe$_{2}$O$_{3}$ nanoparticles}
\end{figure}

Keeping in mind that STS is a highly surface sensitive technique
we carried out optical spectroscopy on nanoparticles of the
$\chi$-Fe$_{2}$O$_{3}$ phase, in order to obtain information
corresponding to "bulk properties". Figure~\ref{fig7} shows the
optical transmission data taken from a thick layer of such
particles. The spectrum displays a strong decrease of transmission
at around 2.0~eV, and a less pronounced decrease at about 1.3~eV.
These energies agree well with the two different gap energies
inferred from STS measurements on such particles. In a straight
forward interpretation we assign the feature at 2 eV to the bulk
gap, the feature at 1.3~eV to a surface gap. This interpretation
will be discussed in more detail below.

\begin{figure}
\includegraphics[width=8.5cm]{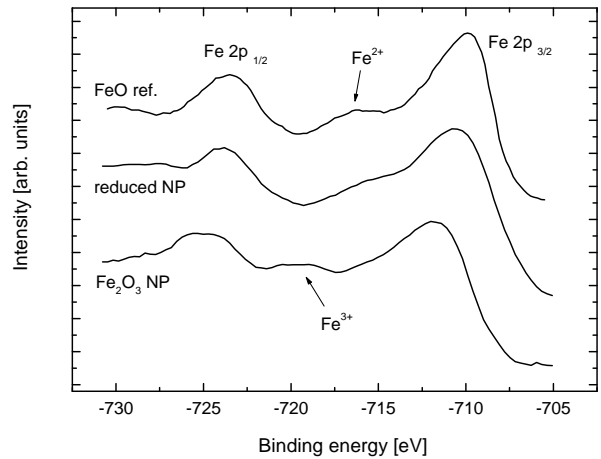}
\caption{\label{fig8}XPS spectra of oxidized and reduced iron
oxide nanoparticles compared to FeO reference data}
\end{figure}

An interesting question concerns the possibility to manipulate the
electronic properties of the nanoparticles. To this end we exposed
$\chi$-Fe$_{2}$O$_{3}$ nanoparticles \textit{in situ} at a
temperature of $\sim$~675~K to a reducing H$_2$/Ar - atmosphere of
1$\times10^{-6}$~ mbar (25\%~H$_2$, 75\%~Ar) for two hours.
Figure~\ref{fig8} compares the Fe~2p spectrum of the nanoparticles
before and after reduction. As is clearly seen, the spectrum of
the reduced sample is shifted to lower binding energies by about
1.2~eV compared to the spectrum before reduction. In addition, a
Fe$^{2+}$ satellite structure develops, marked by an arrow in
Figure~\ref{fig8}.

\begin{figure}
\includegraphics[width=8.5cm]{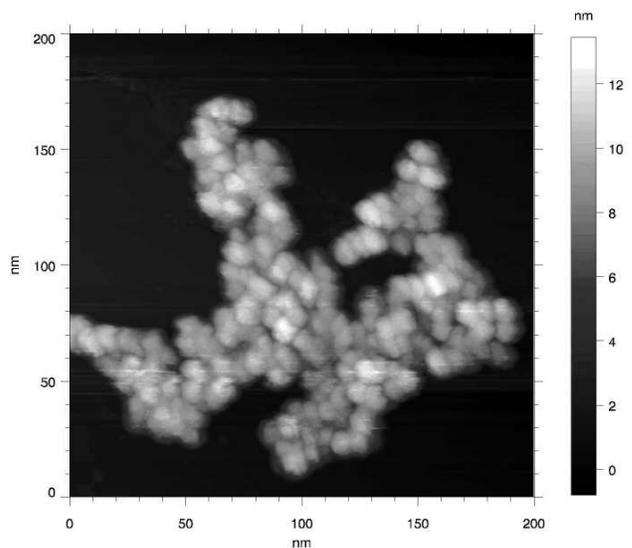}
\caption{\label{fig9}STM image of the reduced
$\chi$-Fe$_{2}$O$_{3}$ nanoparticles}
\end{figure}

\begin{figure}
\includegraphics[width=8.5cm]{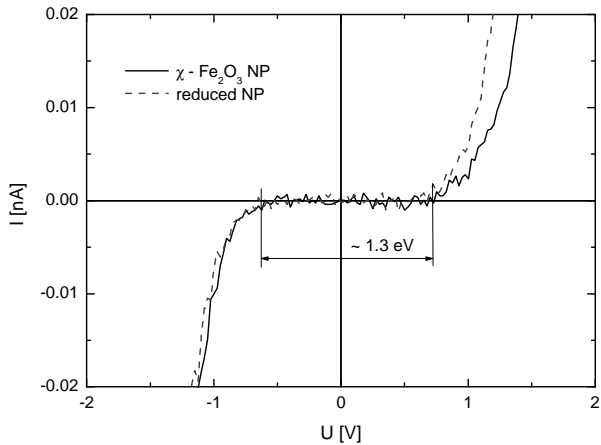}
\caption{\label{fig10}STS spectra of oxidized and reduced iron
oxide nanoparticles}
\end{figure}

STM imaging shows (Figure~\ref{fig9}) that shape and size of the
nanoparticles are not changed by the reduction process. From STS
measurements we infer, within the margins of error,  the same gap
of 1.3~eV ($\pm$~0.2~ eV) as before the reduction
(Figure~\ref{fig10}).

\section{Discussion}

We first address the question of the valence of the iron ions in
the nanoparticles. From the Fe~2p binding energy and the position
and intensity of the corresponding satellite structure as measured
by XPS we conclude that the iron valence state of the oxidized
samples is Fe$^{3+}$. This is in agreement with expectations since
it is well known that ferrihydrites transform into $\alpha$- or
$\gamma$-Fe$_{2}$O$_{3}$ at 573~-~723~K under air\cite{Ironox96}.
Further evidence for an Fe$^{3+}$- state after annealing under air
comes from XRD and TEM\cite{Bremser04}, which show the
nanoparticles to consist of different phases of Fe$_{2}$O$_{3}$,
depending on the oxidation procedure. Bulk Fe$_{2}$O$_{3}$ is an
insulator with a gap of $\Delta$~=~2.0~eV for the $\gamma$-phase
and $\Delta$~=~ 2.2~eV for the $\alpha$-phase. The gap energy
obtained for $\alpha$-phase nanoparticles by STS is consistent
with that for bulk material. The $\gamma$-Fe$_{2}$O$_{3}$ phase
nanoparticles show a band gap of 1.3~eV or 2.0~eV, depending on
the location of the tunnelling tip on the same particle. This
suggests two different types of phases within one nanoparticle.
The fact that bulk $\gamma$-Fe$_{2}$O$_{3}$ has a band gap of
2.0~eV\cite{Ironox96} suggests that it is one of these phases.
From single crystal studies it is known that
$\gamma$-Fe$_{2}$O$_{3}$ crystallizes in an iron deficient
structure similar to the inverse spinel structure of
Fe$_{3}$O$_{4}$ (A[BA]O$_{4}$, A=Fe$^{3+}$, B=Fe$^{2+}$), but in
$\gamma$-Fe$_{2}$O$_{3}$ there are randomly distributed cation
vacancies on the B-site. Such a defect rich structure might be
subject to reconstruction effects, especially at the surface of a
nanoparticle or by impurity atoms like, e.g. carbon introduced by
the original organic shell of the particle. Such a reconstructed
phase can yield a different gap than the bulk. The XPS and FTIR
measurements might be affected by such surface effects in a much
lesser degree because of their lower surface sensitivity. This
will be discussed in more detail below.

The FTIR-spectrum taken from the $\chi$-Fe$_{2}$O$_{3}$ phase
sheds further light on the origin of the 1.3~eV gap. Two
significant features were identified in this spectrum, one of
which coincides with the gap of 1.3~eV found by STS. From the fact
that the slope of the absorption coefficient is much smaller at
1.3~eV than at 2~eV, but that the surface sensitive STS
measurements detect only one gap at 1.3~eV we conclude that the
1.3~eV gap is a property of the surface of the nanoparticle, most
likely due to a surface reconstruction. The bulk of the
nanoparticle then exhibits the gap expected from bulk $\gamma$- or
$\alpha$-Fe$_{2}$O$_{3}$. It is striking that a surface
reconstruction would only occur in the $\gamma$-Fe$_{2}$O$_{3}$
and $\chi$-Fe$_{2}$O$_{3}$ nanoparticles. The reason for this
could be attributed to the defect rich nature of these phases.

The interpretation of the 1.3~eV gap as a surface effect is
further corroborated by our reduction experiments. There are
several studies concerning the reduction of iron oxides. For
example, $\alpha$-Fe$_{2}$O$_{3}$ can be reduced to metallic iron
via a series of intermediate oxides, like Fe$_{3}$O$_{4}$ and
FeO\cite{Rau87}. As shown above we achieved a significant
reduction of $\chi$-Fe$_{2}$O$_{3}$ nanoparticles by annealing in
H$_{2}$/Ar - atmosphere, as evidenced by XPS measurements. The
resulting XPS spectrum is compatible to a FeO reference spectrum
taken from a macroscopic sample. However, a stable
cation-deficient phase of FeO exists at ambient pressure only at
temperatures above 840~K. This non-stoichiometric form of FeO can
be obtained as a metastable phase at room temperature by rapid
quenching\cite{Ironox96}. Although we cannot expect a stable FeO
bulk phase from the applied annealing procedure under ambient
conditions, we cannot exclude that the FeO phase is stabilized
under UHV due to its nanocrystalline form. This conclusion is
supported by former studies on epitaxial films showing that the
stabilization of a FeO phase depends on film
thickness\cite{Weiss02,Koveshnikov00}.

Interestingly, a change of the iron oxidation state under reducing
conditions, as reflected in the XPS measurements, does not affect
the energy gap of the nanoparticles as measured by STS. Taking
into account the probing depth of XPS at the Fe 2p binding energy
we conclude that about 40~\% of the volume of a particle
contributes to the XPS spectrum, while contributions to the STS
measurements originate from a surface layer of the particle, which
amounts to only 10~\% of its volume. Obviously, for nanoparticles
XPS is far more bulk sensitive than STS. Therefore, we attribute
the independence of the gap value upon iron reduction to the fact
that the gap value measured by STS is dominated by the surface
structure of the nanoparticle. This surface structure appears to
be unaffected by reduction. This finding is consistent with our
the conclusion above that the 1.3~eV band gap of
$\chi$-Fe$_{2}$O$_{3}$ is due to surface effects.

In conclusion, only nanoparticles of the thermodynamically stable
$\alpha$-Fe$_{2}$O$_{3}$ phase show a band gap compatible to the
bulk material in surface sensitive STS measurements. The STS
measurements on all other nanoparticle phases investigated
($\gamma$-Fe$_{2}$O$_{3}$, $\chi$-Fe$_{2}$O$_{3}$ and FeO)
demonstrate the existence of a surface band gap of 1.3~eV,
although volume sensitive measurements reveal a band gap
compatible with the bulk material. We, therefore, conclude that
the band gap does not depend on the size of the nanoparticles for
diameters above 8~nm, but that properties, like e.g. the
photocatalytic activity of such particles, are due to the
formation of a surface band gap different from that of the volume.
The existence of such a surface band gap appears to be triggered
by defects within the nanoparticle.

\begin{acknowledgments}
This work was supported by the Deutsche Forschungsgemeinschaft
through SFB~484.
\end{acknowledgments}

\end{document}